\documentstyle[11pt,epsfig,twocolumn]{article}
\title{{\bf Global stability  of systems related to the 
Navier-Stokes equations}}
\author{{\bf Alexander Rauh}\\
{\small  Fachbereich Physik, Carl von Ossietzky Universit\"at,
D-26111 Oldenburg, Germany }}

\date{}

\begin{document}

\maketitle

\newcommand{\beq}{\begin{equation}}                                            
\newcommand{\eeq}{\end{equation}}          
\newcommand{\pa}{\partial}                                                    
\newcommand{\jm}{\jmath}

\begin{abstract}
A generalized Lyapunov method is outlined which predicts global
stability of a broad class of dissipative dynamical systems. 
The method is applied  to the complex Lorenz model and to
the Navier-Stokes equations.
In both cases one finds compact  domains in phase
space which contain the $\omega$ sets of all trajectories, in particular
the fixed points, limit cycles, and strange attractors.
\end{abstract}

\section{Introduction}

In the theory of ordinary differential equations, the method of
 Lyapunov function in general serves to examine the stability
of a fixed point and its domain of attraction, for an overview see e.g.
\cite{lasalle}.
 The method can
be naturally extended to the case where, instead of a fixed point, one is 
interested in the stability of a compact domain which has  finite measure 
in phase space and is invariant under the flow. 

In this 
contribution we essentially illustrate the power of generalized
Lyapunov functions, which are not discussed in the standard textbooks on
dynamical systems theory. In particular we are interested in finding
globally attractive domains 
for  a certain class of nonlinear models. The corresponding systems turn out
to be globally stable in the sense that no trajectory which starts within
a certain  domain can leave it and the trajectories which start outside of the
domain will end in it after sufficiently large times. 
The method presented cannot give details on the nature of the 
attractors contained within an attractive domain. Furthermore, we will
only partially succeed to determine  minimal attractive domains. 
This, on the other hand,
opens  the chance of finding attractive domains in an analytical
way. As a matter of fact, in his  famous paper, Lorenz showed
\cite{lorenz}  that (nonminimal) attractive domains can be found in an 
elementary way by  linear methods provided
the nonlinearities of the dissipative dynamical system are quadratic only 
and do not contribute to the overall energy balance.    

In the next section the method of  generalized Lyapunov functions
will be introduced together with a class of dynamical systems 
as proposed by Lorenz \cite{lorenz}, which allow for quadratic Lyapunov
functions.  As a first example,  the method is applied
to the real Lorenz model according to \cite{sparrow}. 
 We discuss then in the third section
a more detailed application to the five-dimensional or complex Lorenz model
\cite{Fowler}. As compared with  recent work \cite{rahaab} 
where the attractive domain has been successfully minimized to some extent
for parameter values  relevant in infra-red laser physics,
further new results \cite{frank} are presented here. 
In the last section  the generalized Lyapunov method serves to 
prove  the boundedness of the velocity field 
of the incompressible Navier-Stokes equations. This  is a known result and 
was shown in different ways elsewhere, for  the case of periodic
boundary conditions see e.g. section 5.3 in \cite{doering}.

\section{Generalized Lyapunov functions}

Let us consider an autonomous dynamical system 
\beq
\frac{d x}{d t}=f(x); \hspace{0.5cm} f:\,\,\, {\bf R}^n \rightarrow 
{\bf R}^n
\eeq
where the vector field $f$ is supposed to be sufficiently smooth.
Be $L(x)$ a positively definite, sufficiently smooth, scalar function with 
$L(0)=0$ and $L(x) >0$ for $x\neq 0$.
 Furthermore, be $G$ a domain which contains
the point $x=0$, and $\bar{G}$ its complement. Then we call
$L$ a generalized Lyapunov function, if the following properties hold for
a compact domain $G$
\begin{eqnarray}
L(x)  >  0 \hspace{0.5cm} \rm{for} \hspace{0.5cm} x\neq 0 
\hspace{0.5cm} \rm{and}  \hspace{0.5cm} L(0)=0 \\
\frac{d L}{dt} : =  \frac{d}{dt} \, L(x(t,x_0))=\sum^n_{i=1}\frac{\pa L}
{\pa x_i}\, f_i(x)<0  \nonumber \\
 \rm{for}  \hspace{0.3cm} x \in \bar{G}, \\
 grad_x\, L(x)  \neq   0 \hspace{0.3cm} \rm{for}  \hspace{0.3cm} x\neq 0,
\nonumber \\
\rm{and}  \hspace{0.3cm} \rm{Lipschitz}\hspace{0.2cm}
\rm{continuous.}
\end{eqnarray}

The manifolds $L(x)=C$ = constant are closed hypersurfaces, which surround 
the  point $x=0$ and foliate the phase space. To see this, one starts with a
sufficiently small constant $C$ which because of $L(x)>0$ 
is connected  with an ellipsoidal surface. Then one constructs the 
one-dimensional curves normal to the surfaces $L(x)=C$. If $\lambda$
is a suitable curve parameter, the  curves $x(\lambda)$ with
$x:\, {\bf R} \rightarrow {\bf R}^n$
can be defined through the property that the curve tangents  are
parallel to the surface normal in every point. We can thus consider the
curves as trajectories of the following dynamical system
\beq
\frac{d x_i}{d \lambda}=\frac{\pa L(x(\lambda))}{\pa x_i}, \hspace{0.3cm}
i=1,2,...n.
\eeq
Since this system is autonomous and because of (4),
we obtain unique curves $x(\lambda,x_0)$ which do not intersect 
or touch each other;
 $x_0 \in {\bf R}^n$ is an arbitrary initial point which can be chosen, 
for instance, on an ellipsoid close to the origin. In this way we have
constructed a 1-1 map between the points of an arbitrary surface $L=C$
and the points of an ellipsoid close to the origin $x=0$. In other words,
the surfaces $L=C$ are homeomorph to an ellipsoid surrounding $x=0$.
Moreover, any surface with a given constant $C_0$ separates the phase space
into an inner part which contains $x=0$ and an outer part foliated
by the surfaces $C>C_0$.  

At the boundary of $G$ we have points where $dL/dt=0$. We remark that this
 set contains the critical points of the system where
$f(x)=0$, because $dL/dt=\Sigma_i f_i \pa L/\pa x_i=0$.
The surface $L=C^*$ which both
bounds a domain containing the  points with $dL/dt=0$ and is 
minimal with respect to $C$, is called the critical one. Now we are ready 
to draw  conclusions for the trajectories of the system (1). 
If $x_0$ is the initial point
of a trajectory with the property $L(x_0)=C_0>C^*$, then because 
of (3) we have $dL(x_0)/dt<0$. Therefore the trajectory wanders towards inner 
points with  smaller $C$ until the critical surface
with $L(x)=C^*$ is reached. This tells
that the critical surface is attractive from the outside. Simultaneously
there can be no escape of a trajectory which starts inside the domain
bounded by the critical surface. For illustration see Fig.1.
The existence of a generalized Lyapunov function guarantees therefore that
the trajectories of the
dynamical system asymptotically are  confined to the   domain $G^*$
bounded by the critical surface. Clearly, if we find several Lyapunov 
functions with attractive domains $G^*_1, G^*_2,..$, then  
the intersection $G^*=\cap_i G^*_i$ contains the minimal attractive
domain.

\begin{figure}[htbp]
\begin{center}
\epsfig{file=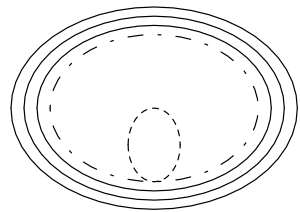,height=4cm}
\end{center}
\caption{ Illustration of a generalized Lyapunov function.
The dashed curve shows the surface $dL/dt=0$, which is tangent to the 
critical  surface $L=C^*$ (dot-dashed). The latter confines a domain of 
attraction. The solid curves refer to surfaces $L=C$ with $C>C^*$
and $dL/dt<0$.}
\end{figure}

As a rather general example Lorenz considered the following
dynamical system, for $i=1,\ldots n$,
\beq
\frac{dx_i}{dt}\equiv \dot{x_i}=\sum_{j,k} a_{ijk} x_j x_k\;-\;
 \sum_j b_{ij} x_j\;+\;c_i
\eeq
with
\beq
\sum_{i,j,k} a_{ijk} x_i x_j x_k\equiv 0   \hspace{0.4cm} \rm{and} 
\hspace{0.4cm} \sum_{i,j} x_i\, b_{ij}\, x_j>0.
\eeq
He proposed the Lyapunov function
\beq
L(x)=\frac{1}{2} (x^2_1+x^2_2+... x^2_n)
\eeq
which gives rise to
\beq
\dot{L}=-\sum_{i,j} x_i \,b_{ij}\,x_j+\sum_i c_i\,x_i.
\eeq
Now, because  the symmetric part of the matrix $b_{ij}$ is positively definite 
(all eigenvalues are positive), $dL/dt<0$ for sufficiently
large $|x|$. Therefore $L$ fulfils all conditions (2)-(4) of a generalized
Lyapunov function.  

As an elementary example we consider the Lorenz model \cite{lorenz}
\beq
\dot{x}=\sigma (y-x); \hspace{0.4cm} \dot{y}=-x z +r x -y;  \hspace{0.4cm}
\dot{z}=x y-b z
\eeq
with $r,\sigma,b>0.$
Sparrow, see Appendix C in \cite{sparrow}, proved the boundedness of this
model with the aid of the following function
\beq
\tilde{L}=r x^2+\sigma y^2+\sigma (z-2 r)^2.
\eeq
After the coordinate shift $x_1:=x$, $x_2:=y$, $x_3:=z-2r$, we obtain
\beq
L(x):=r x^2_1+\sigma x^2_2+\sigma x^2_3
\eeq
and
\beq
\dot{L}=-2 \sigma (r x^2_1+x^2_2+b x^2_3+2 b r x_3)
\eeq
which is negatively definite for sufficiently large distances
$\sqrt{x^2_1+x^2_2+x^2_3}$. Thus $L(x)$ as defined in (12) fulfils
the conditions of a generalized Lyapunov function with the implication
that  a bounded domain exists which attracts all trajectories.

\section{Application to the complex Lorenz model}

The complex Lorenz model reads in standard form \cite{Fowler} 
\begin{eqnarray}
\dot{X} & = & -\sigma X+\sigma Y \nonumber \\ \dot{Y} &
= & -aY+ rX-XZ \label{3} \\ \dot{Z} & = & -bZ+\frac 12(X^{*}Y+XY^{*})
\nonumber \end{eqnarray}
where $X,Y$ and $Z$ are complex variables and real, respectively. Furthermore,
$a=1-ie$, $r=r_1+ir_2$ with real parameters 
$e$, $r_1$, $r_2$, $\sigma $, and $b$. 
In the case of modeling a detuned laser, the 
constants $r_1,r_2$ are related to the pumping rate and to the detuning,
respectively. Furthermore $\sigma =\kappa /\gamma _{\bot }$ and $b=\gamma
_{\Vert }/\gamma _{\bot }$ where $\kappa ,\gamma _{\bot },\gamma _{\Vert }$
denote the relaxation constants of the cavity, of the polarization, and of
the inversion. 
The variable $X$ is proportional to the complex electric field amplitude,
$Y$ is a linear combination of electric field and polarization, which are 
both complex, while $Z$ is related to the so-called population inversion,
for details see e.g. \cite{BA}.
As is well known \cite{Fowler}, this model has nontrivial
stationary solutions only in the so-called laser case
with the parameter constraint $e=-r_2$.

It is convenient to introduce real variables $x_i$, with $%
i=1,...,5$, by $X=x_1+ix_2,Y=x_3+ix_4$ and $Z=x_5$.  The real
version of (\ref{3}) then reads 
\begin{eqnarray}
\displaystyle\dot{x_1} & = & -\sigma x_1+\sigma x_3 \nonumber \\
\displaystyle\dot{x_2} & = & -\sigma x_2+\sigma x_4 \nonumber\\
\displaystyle\dot{x_3}& = & r_1x_1-x_3-r_2x_2-ex_4-x_1x_5 \label{7} \\
\displaystyle\dot{x_4} & = & r_1x_2-x_4+r_2x_1+ex_3-x_2x_5
\nonumber\\
\displaystyle\dot{x_5} & = & -bx_5+x_1x_3+x_2x_4. \nonumber
\end{eqnarray}

In \cite{rahaab} the following Lyapunov function was proposed
\begin{equation}
\label{12}\tilde{L}=D^2(x_1^2+x_2^2)+x_3^2+x_4^2+(x_5-r_1-D^2\sigma )^2 
\end{equation}
which has the Lie derivative 
\begin{eqnarray}
\displaystyle\frac 12\displaystyle\frac{d\tilde{L}}{d t }=
 -\sigma D^2(x_1^2+x_2^2)-x_3^2-x_4^2-
\nonumber \\
r_2(x_2x_3-x_1x_4)-bx_5(x_5-r_1-
D^2\sigma ). \label{13}
\end{eqnarray}
Here $D$ is an arbitrary parameter at our disposition.
The latter expression turns out to be negatively definite for sufficiently
large distances $\sqrt{x^2_1+...x^2_5}$ provided $D$ obeys the condition
\beq
 r^2_2/(4D^2\sigma)<1. 
\eeq
After the coordinate shift $x_5':=x_5-r_1-D^2 \sigma$, the function
$L(x_1,x_2,x_3,x_4,x_5'):=\tilde{L}(x_1,...x_5)$ fulfils all
requirements (2)-(4) of a generalized Lyapunov function. It is thus
proved that also the complex Lorenz model is bounded 
for all parameters, with and without the laser condition $e=-r_2$
\cite{rahaab}.

For quantitative results one determines the ellipsoid $L=C^*$ which
touches the (geometrically different) ellipsoid $dL/dt=0$ from
the outside. This amounts
to a five-dimensional secular problem which in the given case
happens to be feasible analytically. The attractive domain is then
minimized with respect to the parameter $D$ with due attention paid to
the constraint (18). Details can be found in \cite{rahaab}. Numerical
evaluations for physically relevant parameters give upper bounds for
the laser electric field which exceed the maximum values reached 
by asymptotic solutions of (15) by factors of  between 2 and 6. In
extreme cases of transient evolution, the solutions approach
within 20\%  of the upper bounds predicted by the Lyapunov method,
see Fig.4 and 5 in \cite{rahaab}. 

As a remark, we have examined  the more general Lyapunov function 
\cite{frank}
\begin{eqnarray}  
L & = & D^2 (x_1^2 + x_2^2) + x_3^2 + x_4^2 
+ 2\xi\, (x_2 x_3 - x_1 x_4) 
\nonumber \\
&+& (x_5 -\nu)^2 ;\nonumber\\
\nu & = & r_1 +D^2 \sigma + \xi r_2;  
\hspace{0.7cm} \xi^2 < D^2, 
\end{eqnarray}
with the further  disposable parameters $\xi$ in addition to $D$.
$L$ fulfils the properties (2) and (4) as is immediately seen after the
coordinate transformation $x_1 \rightarrow x_1-\xi/D^2\, x_4$,
$x_2 \rightarrow x_2+\xi/D^2\, x_3$,  $x_3 \rightarrow x_3$,
 $x_4 \rightarrow x_4$,  $x_5 \rightarrow x_1-\nu$.
The Lie-derivative is given as
\begin{eqnarray}
\frac{1}{2} \frac{dL}{d t} & =&  -(D^2 \sigma +\xi r_2) (x_1^2 + x_2^2)
-x_3^2 -x_4^2 -
\nonumber\\
&&[r_2 + (\sigma + 1)\xi] (x_2 x_3 - x_1 x_4) \nonumber\\
& \phantom{{=}} & -b x_5^2 +b\nu x_5. 
\end{eqnarray} 
This derivative turns out to be negatively definite for sufficiently
large distances $\sqrt{x^2_1+...x^2_5}$, and thus obeying (3),  provided
\begin{equation}
\frac{[r_2+(\sigma+1)\xi]^2}{D^2 \sigma + \xi r_2} < 4.
\label{formel4}
\end{equation}
The determination of upper bounds of the electric field amplitude
$|X|^2$ is carried out in a similar way as in \cite{rahaab}. The main
challenge consists in the task of simplifying rather involved analytical
expressions for different parameter regions. In a physically
relevant parameter domain  
\beq
r^2_2 \leq \frac{4 \sigma r_1 (2-b) (2 \sigma-b)}{(\sigma+1)^2+(2-b) 
(2 \sigma-b)},
\eeq
with $b < 2$ and $2\sigma > b$,
the following upper bound is found \cite{frank} which is minimized with
respect to the two parameters $D$ and $\xi$
\beq
|X|^2 \leq 4 \sigma r_1\,\, \frac{(\sigma+1)^2}{(\sigma +1)^2+(2-b)
 (2\sigma -b)}.
\eeq
It is smaller than the upper bound $|X|^2$ $ \leq$ $4 \sigma r_1$
as found previously in \cite{rahaab} with one disposable parameter only,
namely $D$.

\section{Application to the Navier-Stokes equations}

We consider the incompressible Navier-Stokes equations (NSE) 
\beq
\rho_0 \left[ \frac{\partial \mbox{\boldmath $v$}}{\partial t} + \left(
\mbox{\boldmath $v$} \cdot \mbox{\boldmath $\nabla$} \right) \mbox{\boldmath
$v$} \right]  =  -\mbox{grad}(p) + \mu \Delta \mbox{\boldmath $v$} + 
\rho_0\mbox{\boldmath $f$}
\label{eq9}
\eeq
in the space domain $\Omega$ which, at least in one dimension, has a finite
diameter so that it can be located between two parallel planes with finite
distance $l$. Because of $\mbox{div}({\bf v})=0$, the density $\rho_0$ is 
constant.
We are not concerned here with weak solutions, which are discussed in 
\cite{rauh}, and assume that the solutions are sufficiently smooth.
As boundary conditions we adopt the no-slip case with 
${\bf v}|_{\pa \Omega}=0$.

In order to obtain a dynamical system of ordinary differential equations,
 we represent the velocity field ${\bf v}$
 in terms of an orthonormalized  system $\Phi_n \in D(\Omega)$,
with $\mbox{div}(\Phi_n)=0$, n=1,2,.., where $D(\Omega)$ denotes the space
of $C^{\infty}$ functions with compact support in $\Omega$. 
We write
\beq
{\bf v}({\bf x},t)=\sum_{n=1,2,...} c_n(t)\Phi_n({\bf x}) \hspace{0.3cm}
\rm{with} \hspace{0.3cm} c_n \in {\bf R}
\eeq
and define the Lyapunov function as follows
\beq
L(c_1,c_2,...):=\rho_0\,\sum_{n=1,2,..}c_n(t)\,c_n(t)=
\rho_0 \int_{\Omega}dV\,{\bf v}\cdot {\bf v}.
\eeq
This function, obviously fulfils  the conditions (2) and (4). To
verify the property (3) we scalarly multiply the NSE (24) with ${\bf v}$
and integrate over the space $\Omega$. On the left hand side we get
\beq
\frac{1}{2} \frac{d L}{dt}=\rho_0 \int_{\Omega} dV\, {\bf v\cdot}
\frac{\pa {\bf v}}{dt}.
\eeq
Because of $\mbox{div}({\bf v})=0$, the cubic term can be transformed into
the surface integral $\int_{\pa \Omega} dS\, v^2 {\bf \hat{n}\cdot v}$ = 0.
The viscosity term, which  is negatively definite, is estimated with the aid
of the Poincar\'{e} inequality \cite{joseph} as follows
\begin{eqnarray}
\int_{\Omega}dV\, {\bf v} \Delta {\bf v}&=&-
\int_{\Omega}dV\,\frac{\pa v_k}{\pa x_i}
\frac{\pa v_k}{\pa x_i} 
\leq - \frac{2}{l^2} \|{\bf v}\|^2 
\nonumber\\
&=& - \frac{2}{l^2}\sum_{n=1,2,..}
c_n c_n
\end{eqnarray}
where $\|{\bf v}\|^2$ = $\int_{\Omega}dV\, {\bf v \cdot v}$.
The pressure term drops out after partial integration. When the last
term with the force density ${\bf f}$ is estimated by the Schwarz
inequality, we obtain
\beq
\frac{1}{2}\frac{d L}{dt} \leq -\mu  \frac{2}{l^2} \|{\bf v}\|^2+
 \rho_0\|{\bf v}\|\,\, \|{\bf f}\|.
\eeq
This proves that $dL/dt<0$ for sufficiently large $\|v\|^2$ = 
$\sum c_n c_n$.
Thus $L$ possesses also the property (3), and as a consequence
$\|{\bf v}\|$ is asymptotically bounded provided the norm of ${\bf f}$
is finite for all times $t$. From $\dot{L}=0$ we obtain as asymptotic
bound 
\beq
\|{\bf v}\| \leq \frac{l^2}{2 \mu}\rho_0 \max_{t>0} \|{\bf f(t)}\|.
\eeq
As a remark, the problem of possible singularities in the solutions 
of the NSE are connected with the space gradient of ${\bf v}$ rather than
to the velocity itself, see \cite{rauh}. The generalized Lyapunov
method is related to  so-called energy methods, see e.g. \cite{joseph} 
and \cite{straughan}.

\section*{Acknowledgements}
The author is indepted to Frank Buss for a critical reading of the 
manuscript.

\end{document}